\documentclass[11pt,a4paper]{article}

\usepackage{amssymb}
\usepackage[dvips]{graphicx}
\usepackage{bm}

\begin{document}

\title{\bf NSVZ-like scheme for the photino mass in softly broken ${\cal N}=1$ SQED regularized by higher derivatives}

\author{
I.V.Nartsev, K.V.Stepanyantz\\
{\small{\em Moscow State University}}, {\small{\em Faculty of Physics, Department  of Theoretical Physics}}\\
{\small{\em 119991, Moscow, Russia}}}

\maketitle

\begin{abstract}
In the case of using the higher derivative regularization we construct the subtraction scheme which gives the NSVZ-like relation for the anomalous dimension of the photino mass in softly broken ${\cal N}=1$ SQED with $N_f$ flavors in all loops. The corresponding renormalization prescription is determined by simple boundary conditions imposed on the renormalization constants. It allows fixing an arbitrariness of choosing finite counterterms in every order of the perturbation theory in such a way that the renormalization group functions defined in terms of the renormalized coupling constant satisfy the NSVZ-like relation.
\end{abstract}

\unitlength=1cm

\section{Introduction}
\hspace*{\parindent}

The NSVZ $\beta$-function is an important equation relating the $\beta$-function and the anomalous dimension of the chiral matter superfields in ${\cal N}=1$ supersymmetric theories \cite{Novikov:1983uc,Jones:1983ip,Novikov:1985rd,Shifman:1986zi}. It gives the exact $\beta$-function in the form of the geometric series for the pure ${\cal N}=1$ supersymmetric Yang--Mills (SYM) theory and can be used for proving the finiteness of ${\cal N}=2$ SYM beyond the one-loop approximation \cite{Shifman:1999mv,Buchbinder:2014wra,Buchbinder:2015eva}. Therefore, it is closely connected to the non-renormalization theorems in theories with extended supersymmetry.
For the ${\cal N}=1$ supersymmetric quantum electrodynamics (SQED) with $N_f$ flavors the NSVZ relation has the form \cite{Vainshtein:1986ja,Shifman:1985fi}

\begin{equation}\label{NSVZ_SQED_Beta_Function}
\widetilde\beta(\alpha) = \frac{\alpha^2 N_f}{\pi}\Big(1-\widetilde\gamma(\alpha)\Big),
\end{equation}

\noindent
where $\alpha$ denotes the renormalized coupling constant and $\widetilde\gamma(\alpha)$ is the anomalous dimension of the chiral matter superfields. The tildes point that the renormalization group functions are defined in terms of the renormalized coupling constant (see Eq. (\ref{RG_Renormalized}) below). The NSVZ-like equation can be also written for theories with softly broken supersymmetry \cite{Hisano:1997ua,Jack:1997pa,Avdeev:1997vx}. It relates the renormalization of the gaugino mass to the renormalization of the rigid theory by the equation \cite{Hisano:1997ua}

\begin{equation}\label{Hisano_Equation}
\frac{\alpha m}{\widetilde\beta(\alpha)} = \mbox{RGI},
\end{equation}

\noindent
where $m$ is the gaugino mass and RGI means that this expression is the renormalization group (RG) invariant. In this paper we will consider softly broken ${\cal N}=1$ SQED with $N_f$ flavors. In this case, by the help of Eq. (\ref{NSVZ_SQED_Beta_Function}), it is possible to present Eq. (\ref{Hisano_Equation}) in the form of the relation between the anomalous dimension of the photino mass and the anomalous dimension of the matter superfields,

\begin{equation}\label{Gamma_M}
\widetilde\gamma_m(\alpha) = \frac{\alpha N_f}{\pi}\Big[1- \frac{d}{d\alpha}\Big(\alpha\widetilde\gamma(\alpha)\Big)\Big].
\end{equation}

\noindent
where $\widetilde\gamma_m(\alpha)$ is the photino mass anomalous dimension. Again using Eq. (\ref{NSVZ_SQED_Beta_Function}), this relation can be equivalently rewritten as

\begin{equation}\label{NSVZ_Photino}
\frac{d}{d\ln\mu}\Big(\frac{m}{\alpha}\Big) = -\frac{m\alpha N_f}{\pi}\cdot \frac{d\widetilde\gamma(\alpha)}{d\alpha}.
\end{equation}

\noindent
This result can be combined with Eq. (\ref{NSVZ_SQED_Beta_Function}) into the NSVZ-like equation

\begin{equation}\label{NSVZ_General}
\frac{d}{d\ln\mu}\Big(\frac{1}{(1+2m\theta^2)\alpha}\Big) = -\frac{N_f}{\pi}\Big[1-\widetilde\gamma\Big((1+2m\theta^2)\alpha\Big)\Big].
\end{equation}

\noindent
This implies that Eqs. (\ref{NSVZ_SQED_Beta_Function}) and (\ref{NSVZ_Photino}) have a similar nature, which can be revealed using the spurion technique \cite{Girardello:1981wz,HelayelNeto:1984iv,Feruglio:1984aq,Scholl:1984hj,Yamada:1994id}. For example, Eq. (\ref{NSVZ_General}) can be obtained by help of the Statement presented in \cite{Avdeev:1997vx}, which have been derived by this method.

Although the NSVZ and NSVZ-like equations are well-known, to obtain them by explicit summing supergraphs is a complicated problem. At present, it has been solved only in the Abelian case by using the higher derivative regularization \cite{Slavnov:1971aw,Slavnov:1972sq} in the supersymmetric version \cite{Krivoshchekov:1978xg,West:1985jx}.\footnote{This regularization also includes inserting the Pauli--Villars determinants for regularizing the one-loop divergences \cite{Slavnov:1977zf,Faddeev:1980be}.} With this regularization the NSVZ relation in the Abelian case is obtained for the RG functions defined in terms of the bare coupling constant in all orders \cite{Stepanyantz:2011jy,Stepanyantz:2014ima}. Note that this result is valid in an arbitrary subtraction scheme, because the RG functions defined in terms of the bare coupling constant are scheme-independent for a fixed regularization \cite{Kataev:2013eta}. NSVZ-like relations for such RG functions have also been proved in all orders for the Adler $D$-function \cite{Adler:1974gd} in ${\cal N}=1$ SQCD \cite{Shifman:2014cya,Shifman:2015doa} and for the anomalous dimension of the photino mass in the softly broken ${\cal N}=1$ SQED \cite{Nartsev:2016nym}. All these NSVZ and NSVZ-like relations appear due to factorization of loop integrals into integrals of (double) total derivatives in the limit of the vanishing external momentum \cite{Soloshenko:2003nc,Smilga:2004zr}.\footnote{In the case of using the dimensional reduction \cite{Siegel:1979wq} this limit is not well-defined. Therefore, the factorization into double total derivatives does not take place in this case \cite{Aleshin:2015qqc}, and the RG functions defined in terms of the bare coupling constant do not satisfy the NSVZ relation \cite{Aleshin:2016rrr}.} For various non-Abelian theories the factorization into double total derivatives has been verified in the lowest loops \cite{Pimenov:2009hv,Stepanyantz_MIAN,Stepanyantz:2011bz,Aleshin:2016yvj,Buchbinder:2014wra,Buchbinder:2015eva}, but has yet not been derived in all orders. However, in this paper we consider the Abelian case, for which the relations

\begin{equation}\label{Bare_Relations}
\beta(\alpha_0) = \frac{\alpha_0^2 N_f}{\pi}\Big(1-\gamma(\alpha_0)\Big);\qquad \frac{d}{d\ln\Lambda}\Big(\frac{m_0}{\alpha_0}\Big) = -\frac{m_0\alpha_0 N_f}{\pi}\cdot \frac{d\gamma(\alpha_0)}{d\alpha_0},
\end{equation}

\noindent
where $\alpha_0$ is the bare coupling constant, have been rigorously proved in all orders for softly broken ${\cal N}=1$ SQED with $N_f$ flavors, regularized by higher derivatives. These scheme-independent equations relate the RG functions defined in terms of the bare coupling constant. Note that the first equation has been verified by an explicit three-loop calculation in \cite{Kazantsev:2014yna}.

However, the RG functions are standardly defined in terms of the renormalized coupling constant. In this case they depend on the subtraction scheme. In particular, it is possible to verify that the NSVZ relation is also scheme dependent\footnote{General equations which describe, how the NSVZ relation is changed under the finite renormalizations are presented in Refs. \cite{Kutasov:2004xu,Kataev:2013csa,Kataev:2014gxa}.}. If (softy broken) supersymmetric theories are regularized by the dimensional reduction, the NSVZ (or NSVZ-like) scheme can be obtained from the $\overline{\mbox{DR}}$-scheme by a finite renormalization, which should be tuned in every order \cite{Jack:1996vg,Jack:1996cn,Jack:1998uj,Jack:1997eh,Jack:1998iy,Jack:1999aj,Harlander:2006xq,Mihaila:2013wma}. For Abelian rigid theories regularized by higher derivatives the NSVZ scheme is obtained by imposing the simple boundary conditions

\begin{equation}\label{Rigid_Boundary_Conditions}
Z_3(\alpha,x_0)=1;\qquad Z(\alpha,x_0)=1
\end{equation}

\noindent
on the renormalization constants of the charge and of the matter superfields, respectively \cite{Kataev:2013eta,Kataev:2013csa,Kataev:2014gxa}. In these equations $x_0$ is a fixed value of $x\equiv \ln \Lambda/\mu$, where $\Lambda$ is a dimensionful parameter of the regularized theory (which acts as an ultraviolet cut-off) and $\mu$ is a normalization point. (The similar conditions for the non-Abelian case are discussed in \cite{Stepanyantz:2016gtk}.) Under the conditions (\ref{Rigid_Boundary_Conditions}) the $\beta$-function and the anomalous dimension of the matter superfields satisfy Eq. (\ref{NSVZ_SQED_Beta_Function}) in all orders.

In this paper we construct the boundary conditions similar to Eq. (\ref{Rigid_Boundary_Conditions}) under which Eq. (\ref{Gamma_M}) (or, equivalently, Eq. (\ref{NSVZ_Photino}) or Eq. (\ref{NSVZ_General})) is valid in all loops for softly broken ${\cal N}=1$ SQED with $N_f$ flavors.

\section{NSVZ-like scheme for softly broken ${\cal N}=1$ SQED}
\hspace*{\parindent}

For rigid ${\cal N}=1$ SQED the RG functions are defined in terms of the bare coupling constant according to the prescription

\begin{equation}\label{Bare_Beta_Gamma}
\beta(\alpha_0) \equiv \frac{d\alpha_0}{d\ln\Lambda}\Big|_{\alpha=\mbox{\scriptsize const}};\qquad \gamma(\alpha_0) \equiv -\frac{d\ln Z}{d\ln\Lambda}\Big|_{\alpha=\mbox{\scriptsize const}},
\end{equation}

\noindent
where $Z$ is the renormalization constant for the matter superfields, $\phi = \sqrt{Z} \phi_R$. The $\beta$-function can be also expressed in terms of the renormalization constant $Z_3 \equiv \alpha/\alpha_0$,

\begin{equation}
\beta(\alpha_0) = - \alpha_0 \frac{d\ln Z_3}{d\ln\Lambda}\Big|_{\alpha=\mbox{\scriptsize const}}.
\end{equation}

\noindent
In the softly broken theory we also define the anomalous dimension of the photino mass

\begin{equation}\label{Bare_Gamma_M_Definition}
\gamma_m(\alpha_0) \equiv \frac{d\ln m_0}{d\ln\Lambda}\Big|_{\alpha,m=\mbox{\scriptsize const}} = - \frac{d\ln Z_m}{d\ln\Lambda}\Big|_{\alpha=\mbox{\scriptsize const}} = \frac{\alpha_0}{m_0} \frac{d}{d\ln\Lambda}\Big(\frac{m_0}{\alpha_0}\Big) + \frac{\beta(\alpha)}{\alpha_0},
\end{equation}

\noindent
where $Z_m\equiv m/m_0$ is the photino mass renormalization constant. To prove its scheme independence, we consider a part of the two-point Green function of the gauge superfield $\bm{V}$ corresponding to the photino mass. In the limit when the external momentum $p$ is mich larger than all masses it can be written as

\begin{equation}
\frac{m_0}{128\pi} \int \frac{d^4p}{(2\pi)^4} d^4\theta\, \Big(\theta^2 D^a \bm{V}(-p,\theta)\,\bar D^2 D_a \bm{V}(p,\theta) + \bar\theta^2 \bar D^{\dot a} \bm{V}(-p,\theta)\, D^2 \bar D_{\dot a} \bm{V}(p,\theta)\Big) d_m^{-1}(\alpha_0,\Lambda/p),
\end{equation}

\noindent
where $d_m$ is a dimensionless function. Then, according to the definition of the renormalization constant $Z_m$, the product $Z_m d_m(\alpha_0(\alpha,\Lambda/\mu),\Lambda/p)$ should be finite in the limit $\Lambda\to \infty$. Differentiating the logarithm of this expression with respect to $\ln\Lambda$, we express $\gamma_m(\alpha_0)$ via the function $d_m(\alpha_0,\Lambda/p)$,

\begin{equation}\label{Gamma_Vs_Dm}
\gamma_m(\alpha_0) = \lim\limits_{p\to 0} \Big(\frac{\partial \ln d_m(\alpha_0,\Lambda/p)}{\partial\alpha_0}\cdot \beta(\alpha_0) - \frac{\partial\ln d_m(\alpha_0,\Lambda/p)}{\partial \ln p} \Big).
\end{equation}

\noindent
Note that the limit $p\to 0$ is needed in order to get rid of the terms proportional to $(p/\Lambda)^{n}$, where $n$ is a positive integer. The function $d_m(\alpha_0,\Lambda/p)$ is evidently scheme-independent, because it is obtained by calculating the effective action before the renormalization. The scheme-independence of the function $\beta(\alpha_0)$ has been proved in \cite{Kataev:2013eta}. Therefore, the right hand side of Eq. (\ref{Gamma_Vs_Dm}) is also scheme-independent. This implies that the anomalous dimension $\gamma_m(\alpha_0)$ (defined in terms of the bare coupling constant) does not depend on the renormalization prescription. However, it depends on the regularization.

The RG functions (\ref{Bare_Beta_Gamma}) and (\ref{Bare_Gamma_M_Definition}) should be distinguished from the corresponding RG functions (standardly) defined in terms of the renormalized coupling constant \cite{Bogolyubov:1980nc},

\begin{equation}\label{RG_Renormalized}
\widetilde\beta(\alpha) \equiv \frac{d\alpha}{d\ln\mu}\Big|_{\alpha_0=\mbox{\scriptsize const}};\qquad \widetilde\gamma(\alpha) \equiv \frac{d\ln Z}{d\ln\mu}\Big|_{\alpha_0=\mbox{\scriptsize const}};\qquad \widetilde\gamma_m(\alpha) \equiv \frac{d\ln m}{d\ln\mu}\Big|_{\alpha_0,m_0=\mbox{\scriptsize const}}.
\end{equation}

The main observation that allowed constructing the NSVZ scheme for the rigid ${\cal N}=1$ SQED regularized by higher derivatives is that the $\beta$-function and the anomalous dimension defined in terms of the bare coupling constant and the ones defined in terms of the renormalized coupling constant coincide, if the boundary conditions (\ref{Rigid_Boundary_Conditions}) are imposed on the renormalization constants \cite{Kataev:2013eta}. Really, according to \cite{Stepanyantz:2011jy,Stepanyantz:2014ima}, the former RG functions satisfy the NSVZ relation independently of the subtraction scheme in all orders in the case of using the higher derivative regularization. Therefore, the latter RG functions also satisfy it under the conditions (\ref{Rigid_Boundary_Conditions}).

Now, let us consider the softly broken theory. In this case we also impose the boundary conditions (\ref{Rigid_Boundary_Conditions}), because the anomalous dimension of the matter superfields enter Eq. (\ref{Gamma_M}). Then the NSVZ relation (\ref{NSVZ_SQED_Beta_Function}) is valid for the RG functions defined in terms of the renormalized coupling constant and

\begin{equation}\label{RG_Equality}
\widetilde\beta(\alpha) = \beta(\alpha_0)\Big|_{\alpha_0 = \alpha};\qquad \widetilde\gamma(\alpha) = \gamma(\alpha_0)\Big|_{\alpha_0 = \alpha}.
\end{equation}

\noindent
According to Ref. \cite{Nartsev:2016nym}, the NSVZ-like relation

\begin{equation}\label{Bare_Gamma_M}
\gamma_m(\alpha_0) = \frac{\alpha_0 N_f}{\pi}\Big[1- \frac{d}{d\alpha_0}\Big(\alpha_0 \gamma(\alpha_0)\Big)\Big]
\end{equation}

\noindent
is valid in all loops for the theory regularized by higher derivatives. Therefore, we need to find the boundary conditions under which

\begin{equation}\label{Gamma_M_Equality}
\widetilde\gamma_m(\alpha) = \gamma_m(\alpha_0)\Big|_{\alpha_0 = \alpha}.
\end{equation}

\noindent
This can be done by repeating the argumentation of Ref. \cite{Kataev:2013eta}. Let us, in addition to Eq. (\ref{Rigid_Boundary_Conditions}), impose the condition

\begin{equation}\label{M_Bondary_Condition}
m(\alpha,x_0)=m_0.
\end{equation}

\noindent
Equivalently, we fix a value $x_0 = \ln\Lambda/\mu$ and require that the renormalization constants satisfy the equations

\begin{equation}\label{Soft_Boundary_Conditions}
Z_3(\alpha,x_0)=1;\qquad Z(\alpha,x_0)=1;\qquad Z_m(\alpha,x_0)=1.
\end{equation}

\noindent
Then, the anomalous dimension of the photino mass defined in terms of the renormalized coupling constant can be presented as

\begin{equation}\label{Calculation_Of_Gamma_M}
\widetilde\gamma_m\left(\alpha(\alpha_0,x)\right) = - \frac{d}{dx}\ln Z_m\left(\alpha(\alpha_0,x),x\right) = - \frac{\partial \ln Z_m(\alpha,x)}{\partial\alpha} \cdot \frac{\partial \alpha(\alpha_0,x)}{\partial x} - \frac{\partial \ln Z_m(\alpha,x)}{\partial x}.
\end{equation}

\noindent
In this equation $d/dx$ denotes the total derivative with respect to $x=\ln\Lambda/\mu$ which acts both on the explicitly written $x$ and on $x$ inside $\alpha$, unlike the partial derivative $\partial/\partial x$ which does not act on $x$ inside $\alpha$. Let us consider Eq. (\ref{Calculation_Of_Gamma_M}) in the point $x_0$. Then, due to the boundary conditions (\ref{Soft_Boundary_Conditions}), we obtain

\begin{equation}
\frac{\partial \ln Z_m(\alpha,x)}{\partial\alpha}\Big|_{x=x_0} = \frac{\partial \ln(1)}{\partial\alpha} = 0,
\end{equation}

\noindent
so that the first term in Eq. (\ref{Calculation_Of_Gamma_M}) vanishes. In the second term $\ln Z_m$ is differentiated with respect to $\ln\Lambda/\mu$ at a fixed value of the renormalized coupling constant $\alpha$, exactly as in Eq. (\ref{Bare_Gamma_M_Definition}) which defines the anomalous dimension $\gamma_m(\alpha_0)$. Moreover, due to the first boundary condition in Eq. (\ref{Soft_Boundary_Conditions}), $\alpha(\alpha_0,x_0)=\alpha_0$. This implies that both definitions of the photino mass anomalous dimension coincide, see Eq. (\ref{Gamma_M_Equality}).

Therefore, if the boundary conditions (\ref{Soft_Boundary_Conditions}) are imposed on the renormalization constants of softly broken ${\cal N}=1$ SQED with $N_f$ flavors regularized by higher derivatives, then the NSVZ-like equation (\ref{Gamma_M}) (and, consequently, Eqs. (\ref{NSVZ_Photino}) and (\ref{NSVZ_General})) is satisfied in all orders. Thus, we have constructed the prescription which gives the NSVZ scheme in all orders of the perturbation theory.

\section{Conclusion}
\hspace*{\parindent}

In this paper we have constructed the scheme in which the anomalous dimension of the photino mass in softly broken ${\cal N}=1$ SQED with $N_f$ flavors satisfies the NSVZ-like relation (\ref{Gamma_M}) in all loops. In this relation all RG functions are defined in the standard way in terms of the renormalized coupling constant and are, therefore, scheme-dependent. That is why the considered NSVZ-like relation is valid only in a certain subtraction scheme. An important ingredient needed for its constructing is the higher derivative regularization. The matter is that with this regularization the RG functions defined in terms of the bare coupling constant satisfy the NSVZ relation and the NSVZ-like relation for the photino mass in all orders independently of the subtraction scheme. In this paper we proved that, if the boundary conditions (\ref{Soft_Boundary_Conditions}) are imposed to the renormalization constants, the RG functions defined in terms of the bare coupling constant coincide with the RG functions defined in terms of the renormalized coupling constant, see Eqs. (\ref{RG_Equality}) and (\ref{Gamma_M_Equality}). Consequently, the latter RG functions satisfy the NSVZ-like relations in all orders for the higher derivative regularization supplemented by the renormalization prescription (\ref{Soft_Boundary_Conditions}). This construction is very similar to the one presented in \cite{Kataev:2013eta,Kataev:2013csa,Kataev:2014gxa} for the rigid theory. However, in the softly broken theory it is also necessary to impose one more boundary condition on the renormalization constant of the photino mass. Finally, let us again emphasize that this construction is valid only in the case of using the higher derivative regularization.

\section*{Acknowledgments}
\hspace*{\parindent}

K.S. is very grateful to A.L.Kataev for valuable discussions.

The work of K.S. was supported by the Russian Foundation for Basic Research, grant No. 14-01-00695.


\end{document}